\documentstyle[multicol,aps,prl,epsfig]{revtex}
\begin{document}

\newcommand{\be}{\begin{equation}}
\newcommand{\ee}{\end{equation}}
\newcommand{\bea}{\begin{eqnarray}}
\newcommand{\eea}{\end{eqnarray}}
\newcommand{\beb}{\begin{eqnarray*}}
\newcommand{\eeb}{\end{eqnarray*}}
\newcommand{\nn}{\nonumber}

\twocolumn[%
 \csname@twocolumnfalse\endcsname
 \hsize\textwidth\columnwidth\hsize

\preprint{SNUTP 00-035; KIAS-P00074}
\draft
\title{Spontaneous phase oscillation induced by inertia and time delay}
\author{H. Hong,$^1$ Gun Sang Jeon,$^{2}$ and M.Y. Choi$^1$}
\address{
$^1$School of Physics and Center for Theoretical Physics, Seoul National University, 
Seoul 151-747, Korea\\
$^2$Center for Strongly Correlated Materials Research, 
Seoul National University, Seoul 151-747, Korea}

\maketitle

\begin{abstract}
We consider a system of coupled oscillators with finite inertia and
time-delayed interaction, and investigate the interplay between inertia
and delay both analytically and numerically.  
The phase velocity of the system is examined; revealed in numerical simulations
is emergence of spontaneous phase oscillation without
external driving, which turns out to be in good agreement with analytical
results derived in the strong-coupling limit. 
Such self-oscillation is found to suppress 
synchronization and its frequency is observed to
decrease with inertia and delay.  We obtain the phase diagram, which
displays oscillatory and stationary phases in the appropriate regions of
the parameters.
\end{abstract}

%\bigskip

]%
%\begin{multicols}{2}
%\pagebreak
%\addtocounter{page}{1}
 
Systems of coupled nonlinear oscillators, 
%have attracted much attention since they 
serving as a prototype 
model for various oscillatory systems in nature,
%Those systems 
have been known to exhibit remarkable phenomena 
of collective synchronization,
which have been observed in a variety of physical, biological, and 
chemical systems~\cite{Winfree}. 
Among them there are systems where finite inertia and time delay
are relevant. 
For example, in a superconducting junction network~\cite{PRB56},
the capacitance plays the role of inertia
%leading to fractional Shapiro steps in the current-voltage characteristics.
while time delay naturally arises in some physical as well as biological
systems, where finite time interval is actually required 
for the transmission of information~\cite{application}.
This has motivated recent studies of 
the effects of finite inertia and of time delay on synchronization: 
In a system with inertia 
suppression of synchronization and emergence of
hysteresis has been observed~\cite{Hong99},
whereas a multitude of coherent 
states with different synchronization frequencies and suppression of the
collective frequency have been reported in systems with delay~\cite{delay}. 
Note, however, that the effects of the two have been considered separately 
and independently of each other. 
The interplay between the two has not been explored
in the most general system with both inertia and time delay.
%For example, the order parameter of the systems with either inertia or delay 
%has been known to be stationary, which yields conventional synchronization, 
%however the dynamics of the order parameter in the systems 
%with both inertia and delay has not been reported.
%For a single linear oscillator it is known that delayed
%restoring force can change the stability of a fixed point and produce 
%oscillation~\cite{MacDonald}; 
For a single oscillator with inertia, it is known that delayed 
restoring force can destabilize a fixed point and produce 
oscillation~\cite{MacDonald};
this raises an interesting question as to how such oscillatory
%behavior can also be induced by the interplay of inertia and 
%delayed interaction in a {\em nonlinear interacting} system with many
%degrees of freedom.  
behavior affects collective synchronization in a 
{\em set of globally coupled oscillators} with inertia and delay. 

This paper is to investigate the interplay between inertia and 
retarded interaction and their combined effects on synchronization. 
We consider a system of globally coupled oscillators, each possessing
finite inertia and interacting with others via
time-delayed interactions. 
To explore the interplay,
%effects on phase dynamics, 
we first examine the temporal behavior of the phase velocity,
and obtain a nonvanishing ac component in the absence of
external periodic driving. 
Such spontaneous oscillation is found to suppress synchronization,
and its frequency observed to diminish with inertia and delay.  
%We also obtain the phase diagram, which
%displays oscillatory and stationary phases in the appropriate regions of
%the parameters.

We begin with the set of equations of motion governing the dynamics of 
$N$ coupled oscillators, the $i$th of which is described by 
its phase $\phi_i$ $(i=1,2,...,N)$:
\be \label{model}
\mu\ddot{\phi_i}(t)+ \dot{\phi_i }(t) 
+ {K\over N}\sum_{j=1}^N \sin[\phi_i (t) -\phi_j (t{-}\tau)]=
 \omega_i ,
\ee
where $\mu$ represents the magnitude of (rotational) inertia 
relative to damping. 
The third term on the left-hand side denotes the 
global coupling of strength $K/N$ between oscillators,
indicating that each oscillator interacts with others only after
the retardation time $\tau$.  
The term $\omega_i$ on the right-hand side denotes 
the intrinsic frequency of the $i$th oscillator,
and is randomly distributed over the 
whole oscillators according to $g(\omega)$, which is 
assumed to be symmetric about $\omega = 0$, and 
concave at $\omega=0$.
%, i.e., $g''(0)<0$.
In the absence of time delay and inertia ($\tau= \mu=0$), Eq.~(\ref{model}) 
exactly reduces to the Kuramoto model~\cite{Kuramoto}, for which
analytical results are available. 
The system with either time delay or inertia (but not both)
also exhibits synchronization for the coupling strength strong enough.
The phases are stationary or monotonic in time
% such as $\phi=\phi_0 +ct$ with constants $\phi_0$ and $c$
~\cite{Hong99,delay},
indicating that the system either relaxes to the 
minimum-energy configuration or
displays constant phase velocity. 

Here we investigate the dynamics of phases 
in the presence of both inertia and delay, 
first by means of numerical simulations.  
For convenience, we take the Gaussian distribution with unit variance 
($\sigma^2 =1$)
for $g(\omega)$ and choose the coupling strength $K = 3$,
which is larger than the known critical values~\cite{Hong99,delay}.
We thus probe the interplay between inertia and delay 
in the synchronized state.
%%%
Equation~(\ref{model}) has been integrated
with discrete time steps of $\delta t =0.01$. 
In computing the order parameter, $N_t = 10^6 $ time steps have been 
used at each run, with the data from the first $9.6\times 10^5$ steps
discarded.
We have varied both $\delta t$ and $N_t$ to verify that the 
steady state has been attained and performed twenty independent runs 
with different initial configurations, over which averages have 
been taken. 
In this manner, we have computed the power spectrum of the phase velocity:
\be
S \equiv \frac{1}{N}\sum_{j=1}^{N}|\psi_j (f)|^2,
\ee
where $\psi_j (f) \equiv \int \dot\phi_j (t) e^{2\pi i f t}dt $ is the 
Fourier component of the phase velocity at frequency $f$.

Figure 1 shows the obtained power spectrum in a system of 
$N=100$ oscillators, for coupling 
strength $K=3$ and time delay $\tau=2$ at various inertia values,
$\mu=0.4,\,0.6$, and $0.8$. 
(We have considered the size $N$ up to $3200$, and confirmed the absence of
appreciable finite-size effects for $N \gtrsim 100$). 
%Open squares, circles, and triangles in Fig.~1 represent 
%the data for $\mu=0.4,\,0.6$, and $0.8$, respectively. 
For small inertia ($\mu \lesssim 0.3$),
the power spectrum possesses only the zero-frequency component.  
Namely, the system displays stationary or monotonic behavior of the phase,
like the one with delay only~\cite{delay}. 
When the inertia is raised beyond $0.3$, on the other hand, 
several peaks develop at nonzero fundamental and harmonic frequencies, 
and the oscillatory phase emerges.
As $\mu$ is increased further,
those peaks shift to the lower frequency sides, 
eventually disappearing for $\mu \gtrsim 1.0$. 
We have examined such self-oscillatory behavior for various values of 
$K$, $\tau$, and $\mu$,
and observed that the oscillation frequency $f_0$ decreases with
both inertia and delay,
as shown in Fig.~2 for $\tau=1.8,\,2.0$, and $2.2$. 

The resulting phase boundaries in the plane of $(\mu, \tau)$, 
discriminating the oscillatory state and the stationary one,
are displayed in Fig.~3, where the coexistence of the two states is also 
identified in the region between solid squares and open ones.
In the coexistence region, there exist a multitude of solutions,
giving stationary or oscillatory behavior depending on the initial conditions. 
To explore this, we examine the behavior of the order parameter 
$\Delta \equiv N^{-1}|\sum_j \exp (i\phi_j)|$ with the inertia.
Figure~4(a) shows the order parameter computed via numerical simulations,
as the inertia is varied between $0.1$ and $1.2$ with the increment of
$\delta\mu=0.01$.
We have first decreased the inertia from 1.2 to 0.1, which yields the step-like 
structure consisting of the solid circles in Fig.~4(a).  
Upon increasing the inertia from given value, however, the order parameter
does not reverse the path along the solid circles; instead 
it exhibits continuous dependence on the inertia, 
eventually following the open circles.
In this manner a multitude of solutions or ``bands'' have been found,
as shown in Fig.~4(a).
Note that there are two kinds in shape, flat bands and bent ones.  
As the behavior of the phase velocity is examined, 
the system is stationary in the flat bands, 
existing for small inertia ($\mu \lesssim 0.3$) and for large
inertia ($\mu \gtrsim 1.1$).
In the bent bands (for $0.3\lesssim \mu \lesssim 0.9$),
on the other hand, the system displays phase oscillations. 
Accordingly, in the range $0.9\lesssim \mu \lesssim 1.1$, where two kinds of 
bands coexist, either the oscillatory phase or the stationary one is expected to 
appear depending on the initial conditions. 
This coexistence range of inertia (for $\tau =2$) is indeed consistent with Fig.~3. 
%The other feature of the order 
%parameter such as multistability has been investigated in detail in Ref.~\cite{multiband}.

To understand analytically these numerical results, we first 
divide the population of the oscillators into two groups: the synchronized 
group $(S)$ and the desynchronized one $(D)$, and 
take the ansatz $\phi_i = \phi_i^0 + \omega_i t$ for $D$ and 
\be \label{trial}
%\phi_i = \phi_i^0 + \sum_{n=1}^{\infty} A_n \sin(n\Omega t + \alpha_n)
\phi_i = \phi_i^0 + A \sin(\Omega t + \alpha)
\ee
for $S$.
%where $\phi_i^0$ corresponds to the dc component of the phase velocity and 
%the ac component (in the synchronized state) 
%is characterized by the amplitude $A$, the frequency
%$\Omega$, and the initial phase $\alpha$~\cite{harmonics}. 
%%$\Omega$, and the initial phase $\alpha_n$~\cite{harmonics}. 
%For the $D$ group we can employ the expression 
%$\phi_i = \phi_i^0 + \omega_i t $. 
%Since the basin of attraction for nonzero (constant) velocity 
%($\dot\phi_i^0 \neq 0$) is much smaller than that of zero velocity, 
In view of the symmetry of $g(\omega)$, we consider the dc component 
$\phi_i^0$ to be constant.
%For the phase of oscillator belongs to $D$ group we take 
%$\phi_i = \phi_i^0 + \omega_i t $, where $\omega_i$ represents the 
%intrinsic frequency of the oscillator.
Results of numerical simulations also manifest that near the 
phase boundary the ac component is much smaller than the dc component,
allowing the expansion in terms of the amplitude $A$. 
Upon substituting Eq.~(\ref{trial}) into Eq.~(\ref{model}), 
we thus obtain for the oscillators in $S$, to the order of $A^2$, 
\bea \label{expan}
\omega_i 
%&=& - A_1 \mu \Omega^2 \sin (\Omega t + \alpha_1 )
%+ A_1 \Omega \cos(\Omega t + \alpha_1 ) \nonumber\\
%&&+ {K \over N} \sum_{j=1}^{N} \sin\left[\phi_i^0 - \phi_j^0 
%+ 2 A_1 \sin{\Omega\tau \over 2}
%\cos\left(\Omega t  + \alpha_1- {\Omega \tau \over 2}\right)\right]\nonumber \\
&&= {K \over N} {\sum_{j}}^{(S)} \sin(\phi_i^0 {-} \phi_j^0 ) 
\left(1 - A^2 \sin^2 {\Omega \tau \over 2}\right)\nonumber \\
&&  - \left[\mu \Omega^2 - {2K \over N} 
{\sum_j}^{(S)}\cos(\phi_i^0 {-} \phi_j^0 ) \sin^2 {\Omega\tau \over 2} \right] 
A \sin (\Omega t {+} \alpha ) \nonumber \\
&& + \left[\Omega + {K \over N} {\sum_j}^{(S)}\cos(\phi_i^0 {-} \phi_j^0 )
\sin \Omega\tau
\right] A \cos (\Omega t {+} \alpha ), 
\eea
where the summations run over $S$. 
Oscillators in $D$, the fraction of which over the 
whole is negligibly small, $O(\mbox{e}^{-(K\Delta)^2 /2\sigma^2})$, 
turn out not to contribute to the harmonics in $\Omega$. 
%in the large coupling strength, 
% with different frequency from $\Omega$. 
%Further the fraction of the oscillators belong to $D$ group over the whole oscillators
%is very small $O(\mbox{e}^{(K\Delta)^2 /2\sigma}/K\Delta )$, thus 
%it allows us to write as Eq.~(\ref{expan}).
%($\phi_i^0 \simeq \phi_j^0 $) 
A similar expansion for $D$ leads to trivial equations.     

The zeroth harmonics in Eq.~(\ref{expan}) leads to
\be \label{zeroth}
{1 \over N} {\sum_j}^{(S)} \sin(\phi_i^{0} - \phi_j^{0}) = 
%{ \omega_i \over\displaystyle 1 - A_1^2 \sin^2 (\Omega\tau / 2)},
\frac{\omega_i}{K}\left[1+A^2 \sin^2 (\Omega\tau/2) \right]
\ee
or, in terms of the order parameter,
\be \label{deltazero}
\Delta^2 = 1 - 
%\frac{\sigma^2}{K^2 [ 1 - A_1^2 \sin^2 (\Omega\tau / 2)]^2 } 
\frac{\sigma^2}{K^2 }\left[ 1 + 2A^2 \sin^2 (\Omega\tau / 2)\right] 
%\le 1 - \frac{\sigma^2}{K^2}, 
\ee
where $\sigma^2$ is the variance of the distribution $g(\omega)$
and the thermodynamic limit $(N \rightarrow \infty)$ together with the 
strong coupling limit $(K \gg K_c )$ has been taken. 
It is shown in Eq.~(\ref{deltazero}) that 
the appearance of oscillation $(A \neq 0)$ reduces $\Delta$, 
suppressing synchronization. 
%Such phase oscillation phenomena has been observed on the bent band of the 
%order parameter, which has been studied in Ref.~\onlinecite{multiband}. 

We next consider the first harmonics in Eq.~(\ref{expan}), and
obtain~\cite{footnote}
%~\footnote{We replaced the quantities $\sum_j \cos(\phi_i^0 -
%\phi_j^0)/N$ with their mean value $\Delta^2$.
%Indeed the quantities fluctuate with $i$ by the order of ${\cal
%O}(\sigma^2/K^2)$, implying that the ansatz in Eq.~(\ref{trial})
%is exact only for $K \rightarrow \infty$.
%Nevertheless, the smallness of the fluctuation as well as the good 
%agreement between the analytical and the
%numerical results guarantees that the ansatz still provides a good
%approximation for large but finite $K$.}
\bea \label{firstharmonics}
K \Delta^2 (1- \cos \Omega\tau) &=& \mu \Omega^2, \nonumber\\
K \Delta^2\sin \Omega\tau \label{eqDelta0}
&=& - \Omega ,
\eea
which yields only the trivial solution $\Omega =0$ 
in the absence of either inertia or delay.
It is thus concluded that the (sustained) spontaneous oscillation requires both 
inertia and delay. 
Elimination of $\Delta$ in Eq.~(\ref{firstharmonics}) leads to
the equation for the oscillating frequency $\Omega$:
\be \label{eqOmega}
\Omega = - {1 \over \mu} \tan {\Omega \tau \over 2},
\ee
the nonzero solution $(\Omega\neq 0)$ of which may be searched 
in the range $\pi < \Omega \tau < 2\pi$.
Equation~(\ref{firstharmonics}) together with Eq.~(\ref{eqOmega}) 
gives $\Delta^2 $ in the form
\be  \label{delta}
\Delta^2 = {1\over 2K\mu} (1 + \mu^2 \Omega^2 ).
\ee
%\bea 
%(\Delta^0)^2 &=& - {\Omega \over K \sin \Omega \tau} \nonumber\\
%&=& - {\Omega (1 + \tan^2 (\Omega\tau/2))\over 2K 
%\tan(\Omega\tau/2)} \nonumber\\
%&=& {1\over 2K\mu} (1 + \mu^2 \Omega^2 ),
%\eea
Note in Eq.~(\ref{deltazero}) that 
$\Delta$ approaches unity as 
$\sigma / K$ is reduced to zero ($\sigma/K\rightarrow 0$).
In particular Eq.~(\ref{deltazero}) gives the range
%\be \label{constraint}
$
0 \le \Delta ^2 \le 1 - \sigma^2 /K^2,
$
%\ee
where the upper bound of $\Delta ^2$ determines 
the boundary between stationary and oscillatory states.
Accordingly, on the boundary Eq.~(\ref{delta}) reads
\be
\Omega = \sqrt{ {2K\over \mu} 
\left( 1- {\sigma^2 \over K^2}\right) -1 },
\ee
which in turn yields the boundary 
\begin{eqnarray} \label{taurelation}
\tau &=& 2 \left[{2K\over \mu} 
\left( 1- {\sigma^2 \over K^2}\right) - {1\over\mu^2} 
\right]^{-1/2} \nonumber\\
&\times& \left[\pi - \arctan \sqrt{ {2K\mu} 
\left( 1- {\sigma^2 \over K^2}\right) -1 } \right] .
\end{eqnarray}
It can be observed that the oscillatory state appears 
only for $\mu \ge \mu_c \equiv 
\left[ 2K ( 1 - \sigma^2/K^2 )\right]^{-1}$.
The critical value $\mu_c$ decreases, allowing larger 
regions of the oscillatory state,
as the coupling strength $K$ is raised
and as the variance $\sigma^2$ is reduced.
The order parameter $\Delta$ obtained from Eq.~(\ref{firstharmonics}) 
for $K=3$ and $\tau=2$ is plotted by the broken curve in Fig.~4 (b), manifesting 
good agreement with the numerical result.
%in the appropriate regime of $\mu$ and $\tau$ has been observed to be 
%in good agreement with a numerical one [see Fig.~4 (b)].
The tiny discrepancy of $O(10^{-2})$ between the two curves 
can be attributed to the truncation of higher-order terms in Eq.~(\ref{expan}). 

Solving Eq.~(\ref{eqOmega}) in the appropriate regime,
we obtain the oscillation frequency as a function of inertia and delay. 
Figure 5(a) shows the behavior of the fundamental frequency $\Omega/2\pi$ 
on the plane of inertia and delay.  
For comparison, the dependence on inertia for given amount of delay
$\tau=1.8,\, 2.0$, and $2.2$ is shown in Fig.~5(b),
together with the corresponding numerical data (previously plotted in Fig.~2).
Note the good agreement between the analytical results, obtained from 
Eq.~(\ref{eqOmega}), and those from numerical simulations. 
It is manifested that the oscillation frequency in general decreases
with the amount of delay and inertia.
%The behavior of the amplitude of peak with inertia for $K=3.0$ and 
%$\tau=2.0$ is shown in Fig.~3, revealing the extraordinary 
%inertia-delay-induced resonance-like behavior, which shows the 
%qualitative agreement with the analytical results described by 
%Eq.~(\ref{A12}).
%Such amplification of the peak in the system without external 
%periodic driving, 
%thus there is no frequency scale, is very surprising. 

The phase diagram in the three-dimensional space of $(\mu, \tau, K)$
is displayed in Fig.~6(a), where the boundary surface separates 
the oscillatory state from the stationary one. 
Figure 6(b) exhibits the phase boundaries on the plane of inertia and delay 
for several values of the coupling strength.
Also shown are the data obtained from numerical simulations for $K=3$ 
(see Fig.~3), again demonstrating good agreement of the analytical
results based on Eq.~(\ref{taurelation}) with the simulation results. 
It is further observed that the region of the oscillatory state 
grows with the coupling strength.
%Here we note that there is coexistent region of oscillatory and stationary 
%phases owing to the hysteretic behavior, which is further study. 
Note, however, that the analytical approach, focusing on 
the existence of a nonzero solution for $\Omega$, 
is not able to discern the coexistence of oscillatory and stationary 
phases from the oscillatory phase.
Thus the numerical data represented by solid squares in Fig.~3 have 
no counterpart in the analytical results shown in Fig.~6.

In summary, we have studied the synchronization phenomena in 
a system of coupled oscillators, each possessing
finite inertia and interacting with others via
time-delayed interactions. 
The interplay between inertia and time delay
has been investigated in the temporal behavior of the phase velocity,
which reveals the emergence of (sustained) spontaneous oscillation in the absence of
external periodic driving. 
We have also obtained the phase diagram, which
displays oscillatory and stationary states in the appropriate regions of
the three-dimensional space consisting of the inertia, delay, and coupling strength. 

We acknowledge the hospitality from Korea Institute for 
Advanced Study, where part of this work was performed, 
and the partial support from the BK21 Project.
%Korea Research Foundation through Grant 2000-015-DP0138.

\begin{figure}
\vspace*{5.0cm}
\includegraphics{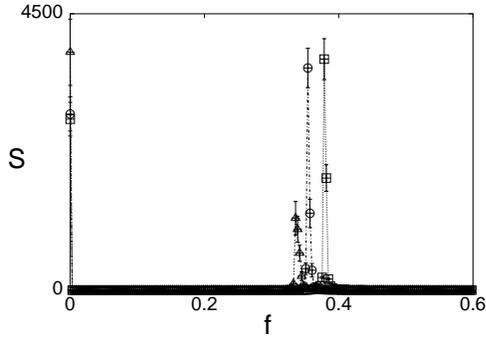}
\vspace{0.5cm}
\caption{Power spectrum of the phase velocity (in arbitrary units) 
for coupling strength $K=3$, time delay $\tau=2$, and inertia 
$\mu=0.4$ (squares), $0.6$ (circles), $0.8$ (triangles). 
Only the fundamental peaks are displayed.
Error bars have been estimated by the standard 
deviation and lines are merely guides to the eye.}
\end{figure}

\begin{figure}
\vspace*{4.2cm}
\includegraphics{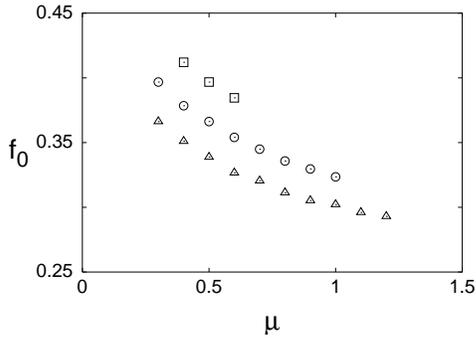}
\vspace{0.5cm}
\caption{Oscillation frequency $f_0$ 
for $K=3$ is shown to decrease with inertia $\mu$ 
at various amounts of delay: $\tau=1.8$ (squares), 
$2.0$ (circles), and $2.2$ (triangles).  The sizes of the error bars 
estimated by the standard deviation are about the same as those of the 
symbols.}
\end{figure}

\begin{figure}
\vspace*{5.0cm}
%\special{psfile=fig3.ps vscale=60 hscale=60 voffset=-325 hoffset=-65}
\includegraphics{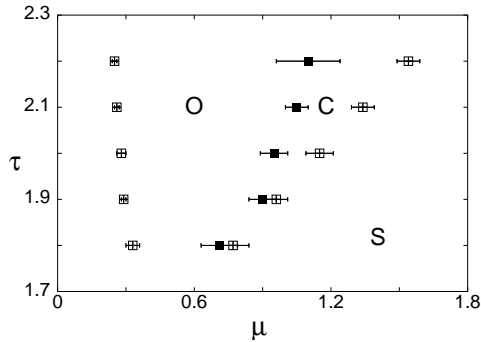}
\vspace{0.5cm}
\caption{Phase boundaries between the oscillatory phase (O) and the
stationary one (S) are exhibited in the plane of $(\mu, \tau)$ for $K=3$. 
Also displayed is the boundary separating the coexistence region (C).
The data represented by open and solid squares have been obtained from 
numerical simulations with five different configurations; 
error bars have been estimated by the range of the obtained values.
}
\end{figure}

\begin{figure}
\vspace*{7.0cm}
%\special{psfile=fig4.ps vscale=45 hscale=45 voffset=-225 hoffset=-20}
\includegraphics{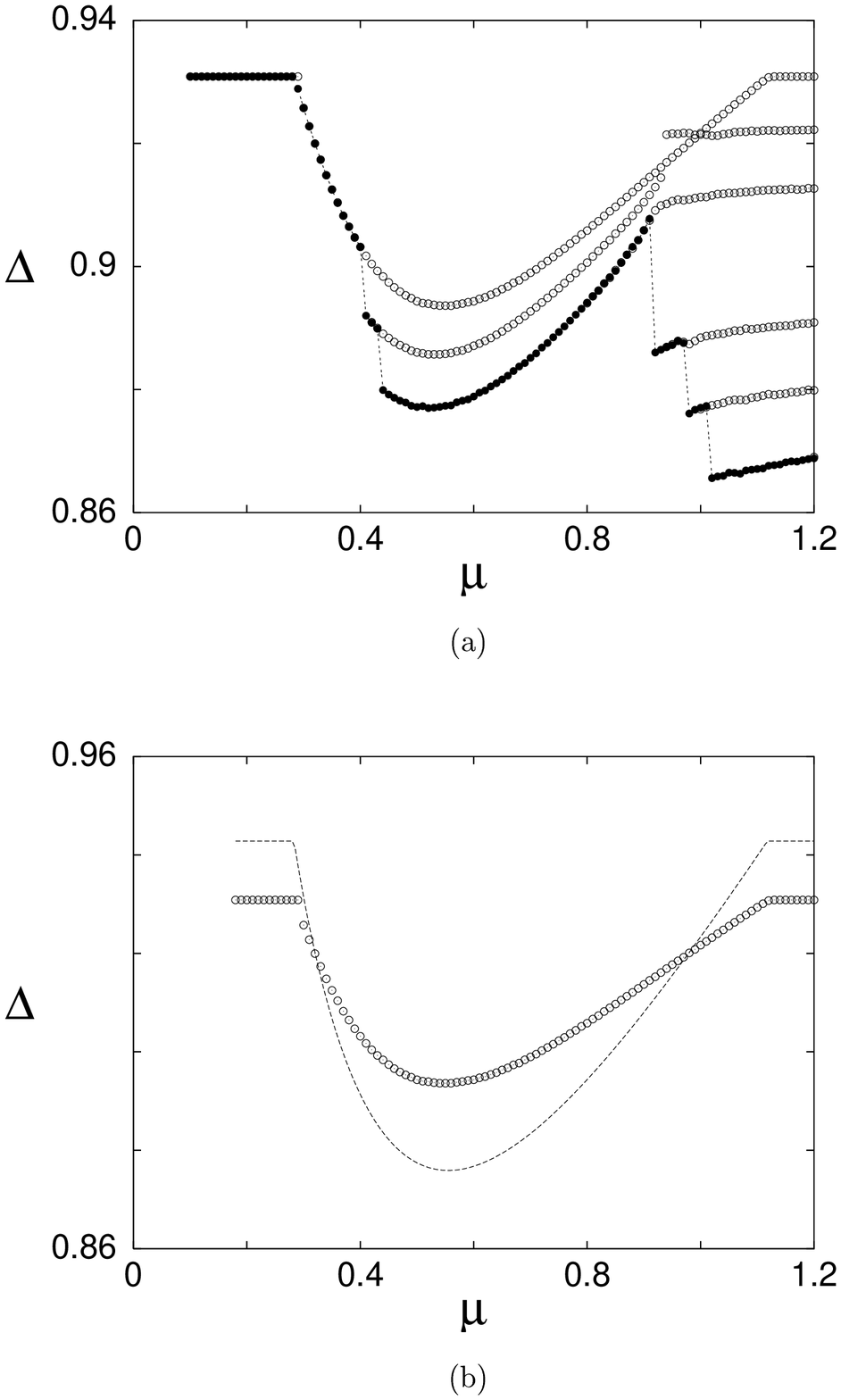}
\vspace{3.0cm}
\caption{(a) Behavior of the order parameter with inertia for 
$K=3$ and $\tau=2$, obtained numerically. (b) Comparison with the 
analytical results (broken curve).
}
\end{figure}

\begin{figure}
\vspace*{7.7cm}
\includegraphics{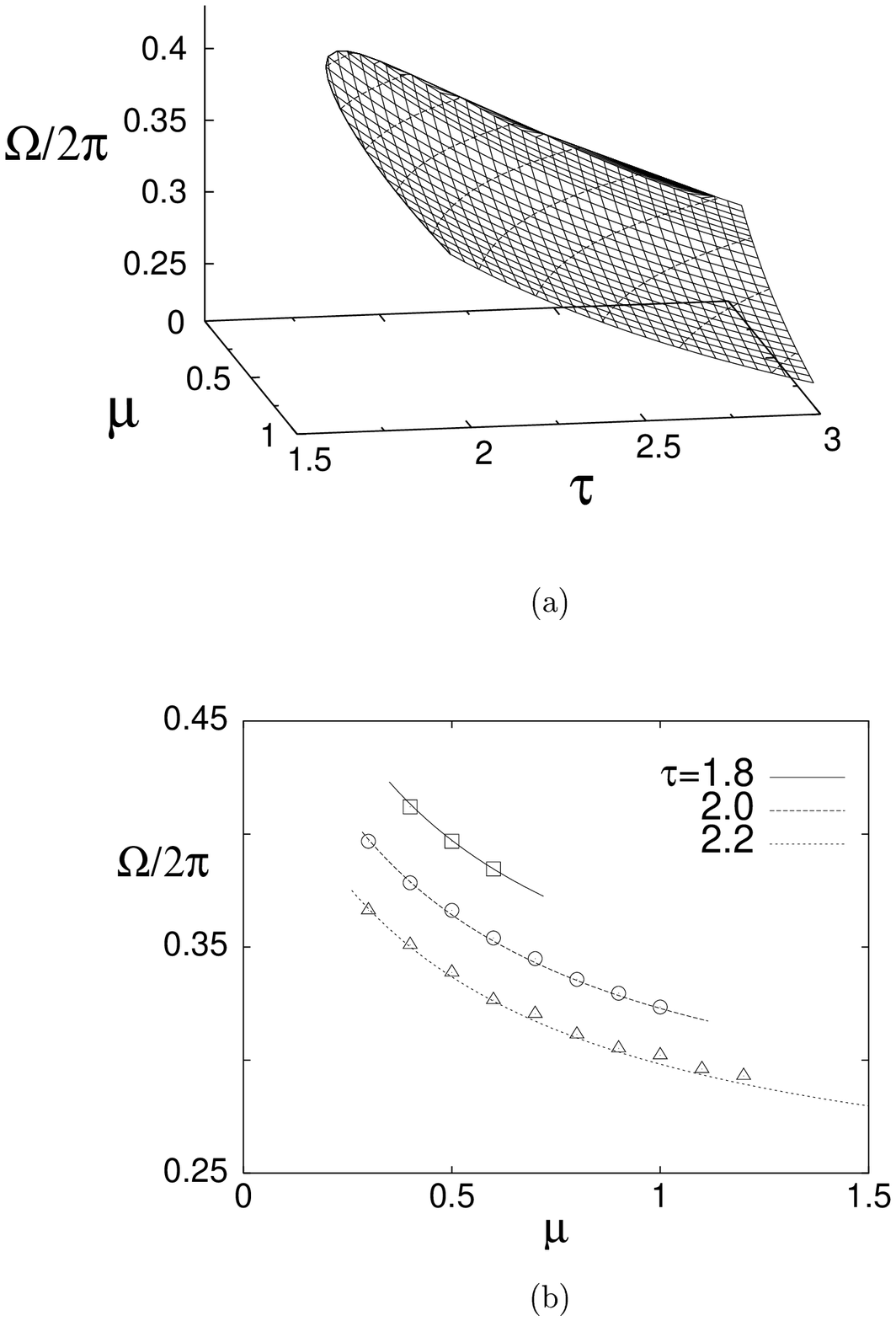}
\vspace{1.7cm}
\caption{Behavior of the fundamental frequency $\Omega/2\pi$ for $K=3$ 
(a) in the $(\mu, \tau)$ plane; (b) versus $\mu$ for $\tau=1.8$, 
$2.0$, and $2.2$, obtained analytically. 
The data from numerical simulations, shown in Fig.~2, are also plotted 
for comparison.
}
\end{figure}

\begin{figure}
\vspace*{7.7cm}
\includegraphics{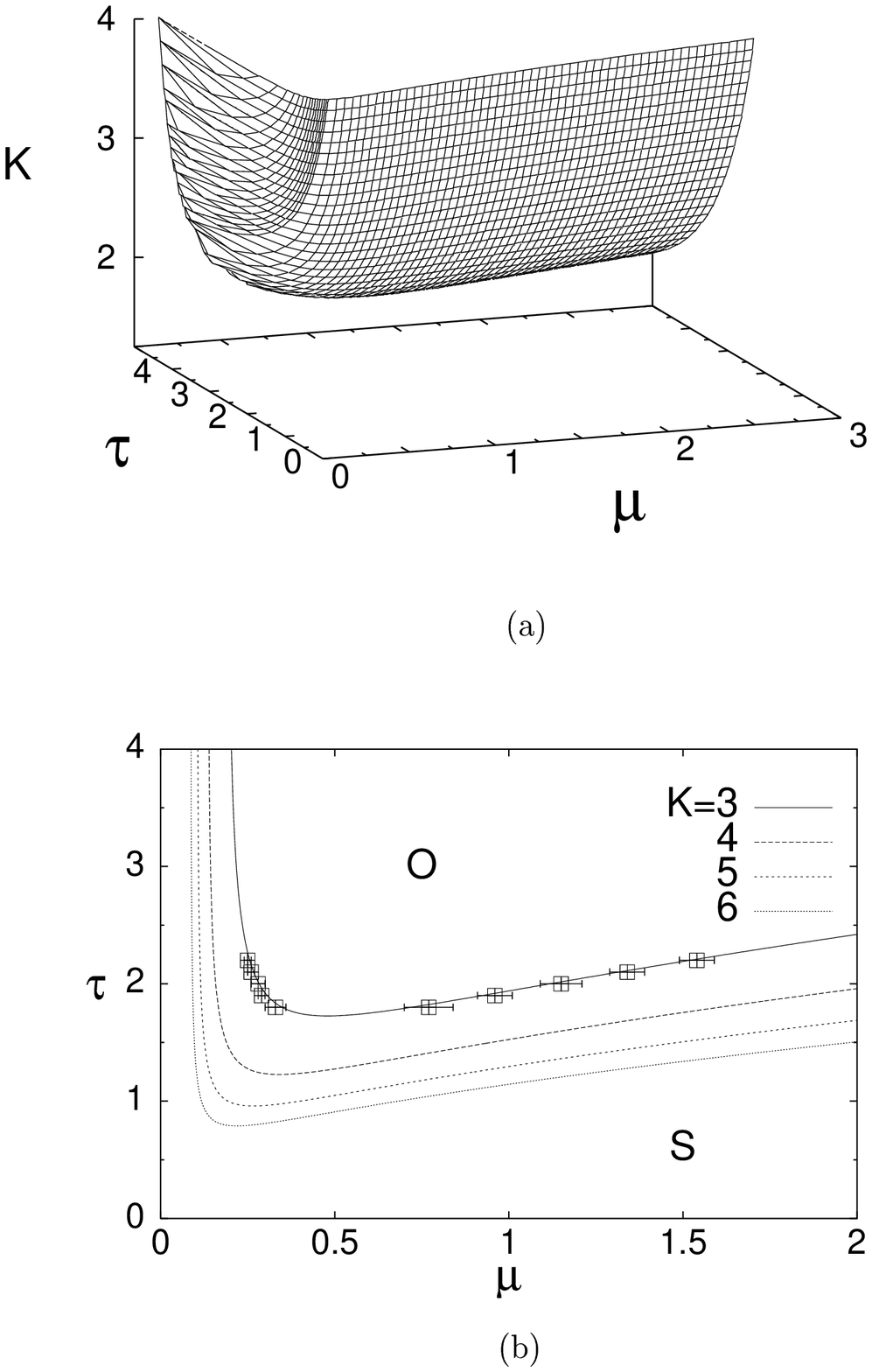}
\vspace{1.3cm}
\caption{Phase diagram 
(a) in the three-dimensional $(\mu, \tau, K)$ space; 
(b) in the $(\mu, \tau)$ plane for several values of the coupling strength,
obtained analytically. 
The data from numerical simulations for $K=3$, shown in Fig.~3,
are also plotted for comparison.
}
\end{figure}

%\end{multicols}


\begin{references}

\bibitem{Winfree}
For a list of references, see A.T. Winfree, {\it The Geometry of Biological Time} 
(Springer-Verlag, New York, 1980);
Y. Kuramoto, {\it Chemical Oscillations, Waves, and Turbulence}
(Springer-Verlag, Berlin, 1984).

\bibitem{PRB56}
K. Park and M.Y. Choi, Phys. Rev. B {\bf 56}, 387 (1997).

\bibitem{application}
D.V. Ramana Reddy, A. Sen, and G.L. Johnston, Phys. Rev. Lett. {\bf 85}, 3381 (2000).

\bibitem{Hong99}
H.-A. Tanaka, A.J. Lichtenberg, and S. Oishi, Phys. Rev. Lett. {\bf 78}, 2104 (1997); 
H. Hong, M.Y. Choi, J. Yi, and K.-S. Soh, Phys. Rev. E {\bf 59}, 353 (1999); 
H. Hong, M.Y. Choi, B.-G. Yoon, K. Park, and K.-S. Soh, J. Phys. A {\bf 32}, L9 (1999); 
J.A. Acebron, L.L. Bonilla, and R. Spigler, Phys. Rev. E {\bf 62}, 3437 (2000); 
H. Hong and M.Y. Choi, {\it ibid}. {\bf 62}, 6462 (2000).

\bibitem{delay}
E. Niebur, H.G. Schuster, and D.M. Kammen, Phys. Rev. Lett. {\bf 67}, 2753 (1991); 
Y. Nakamura, F. Tominaga, and T. Munakata, Phys. Rev. E {\bf 49}, 4849 (1994); 
S. Kim, S.H. Park, and C.S. Ryu, Phys. Rev. Lett. {\bf 79}, 2911 (1997); 
M.K.S. Yeung and S.H. Strogatz, Phys. Rev. Lett. {\bf 82}, 648 (1999);
M.Y. Choi, H. J. Kim, D. Kim, and H. Hong, Phys. Rev. E {\bf 61}, 371 (2000).
%See also

%\bibitem{Periodicsync}
%M. Y. Choi, Y. W. Kim, and D. C. Hong, Phys. Rev. E {\bf 49}, 3825 (1994).
\bibitem{MacDonald}
See, e.g., N.\ MacDonald, {\it Biological Delay Systems: Linear Stability
Theory} (Cambridge University Press, Cambridge, 1989).

\bibitem{Kuramoto}
%Y. Kuramoto, in 
%{\it Proceedings of the International Symposium on Mathematical Problems in 
%Theoretical Physics},
%edited by H. Araki (Springer-Verlag, New York, 1975);
Y. Kuramoto and I. Nishikawa, J. Stat. Phys. {\bf 49}, 569 (1987);
H. Daido, Prog. Theor. Phys. {\bf 77}, 622 (1987); Phys. Rev. Lett. {\bf 68},
1073 (1992).

%\bibitem{multiband}
%H. Hong, G.S. Jeon, and M.Y. Choi, preprint No. SNUTP 01-006. 
%
\bibitem{harmonics}
Here higher harmonics have much smaller amplitudes and are disregarded.

\bibitem{footnote}
We replace the quantity $N^{-1} \sum_j^{(S)} \cos(\phi_i^0 -\phi_j^0)$ 
by its mean value $\Delta^2$.
To be precise, it varies with $i$ in $O(\sigma^2/K^2)$; thus 
the ansatz in Eq.~(\ref{trial})
is exact in the limit $K \rightarrow \infty$.
Such rather small fluctuations as well as the good 
agreement between the analytical and numerical results 
indicate that the ansatz still provides a good
approximation for large but finite $K$.
%\bibitem{basin}
%We note that the velocity ${\dot\phi_i}^0$ is zero in the absence of delay  
%since the average of natural frequencies is set to zero. 

%\bibitem{order}
%We here focus our attention on the case of the large coupling 
%strength ($K\Delta \rightarrow \infty$), which yields 
%$\mbox{e}^{-(K\Delta)^2 /2\sigma^2}/K\Delta \rightarrow 0$. 

\end{references}
\end{document}